\documentclass{webofc}
\usepackage[varg]{txfonts}   
\woctitle{MPGD2015}
\begin{document}
\title{Status of COMPASS RICH-1 Upgrade with MPGD-based Photon Detectors}
%
%
\newcommand*\samethanks[1][\value{footnote}]{\footnotemark[#1]}
\author{M. Alexeev\inst{1}\fnsep\thanks{On leave from JINR, Dubna, Russia.}\and
        R Birsa\inst{2} \and
        F. Bradamante\inst{2,3}\and
        A. Bressan\inst{2,3} \and
				M. Chiosso\inst{1,4} \and
				P. Ciliberti\inst{2,3} \and
				S. Dalla Torre\inst{2} \and
				S. Dasgupta\inst{2,3} \and
				O. Denisov\inst{1} \and
				M. Finger\inst{5,6} \and
				M. Finger Jr\inst{5,6} \and
				H. Fishcher\inst{7} \and
				B. Gobbo\inst{2} \and
				M. Gregori\inst{2} \and
				G. Hamar\inst{2} \and
				F. Herrmann\inst{7} \and
				K. K{\"o}nigsmann\inst{7} \and
				S. Levorato\inst{2} \and
				A. Maggiora\inst{1} \and
				N. Makke\inst{2,3} \and
				A. Martin\inst{2,3} \and
				G. Menon\inst{2} \and
				J. Novy\inst{5,6} \and
				D. Panzieri\inst{1,8} \and
				F. A. Pereira\inst{9} \and
				C. A. Santos\inst{2} \and
				G. Sbrizzai\inst{2,3} \and
				P. Schiavon\inst{2,3} \and
				S. Schopferer\inst{7} \and
				M. Slunechka\inst{5,6} \and
				K. Steiger\inst{2,10}\fnsep\thanks{Their present address is Technical University
				of Liberec, Liberec, Czech Republic and Institute of Plasma Physics,
				Academy of Sciences of the Czech Republic, Turnov, Czech Republic.} \and
				L. Steiger\inst{2,10}\fnsep\samethanks \and
				M. Sulc\inst{10} \and
				F. Tessarotto \inst{2}\fnsep\thanks{corresponding author} \and
				J. F. C. A. Veloso\inst{9} 
}

\institute{INFN Sezione di Torino, Torino, Italy
\and
					 INFN Sezione di Trieste, Trieste, Italy
\and
				   University of Trieste, Trieste, Italy
\and
           University of Torino, Torino, Italy
\and
           Charles University, Prague, Czech Republic
\and
					 JINR, Dubna, Russia
\and
           Universit{\"at} Freiburg, Physikalisches Institut, Freiburg, Germany
\and
           University of East Piemonte, Alessandria, Italy
\and
           Physics Department, University of Aveiro, Aveiro, Portugal
\and
           Technical University of Liberec, Liberec, Czech Republic
          }

\abstract{%
A Set of new MPGD-based Photon Detectors is being built for the upgrade of
COMPASS RICH-1. The detectors cover a total active area of 1.4 m$^2$ and are
based on a hybrid architecture consisting of two THGEM layers and a Micromegas.
A CsI film on one THGEM acts as a reflective photocathode.
The characteristics of the detector, the production of the components and
their validation tests are described in detail.
}
\maketitle
\section{The COMPASS RICH-1 Upgrade}
\label{intro}
The COMPASS Experiment at CERN SPS has recently started a new set of
measurements~\cite{compass2} imposing strict requirements in terms of rate
capability, efficiency, and stability of the detector performance:
several parts of the apparatus are being upgraded to meet these requirements,
including
the COMPASS RICH-1 detector~\cite{rich-1}, which provides $\pi$-K separation
from 3 to 55~GeV/$c$ over $\pm$200~mrad angular acceptance, at high rates.
\par
\begin{figure}
\centering
\includegraphics[width=6cm,clip]{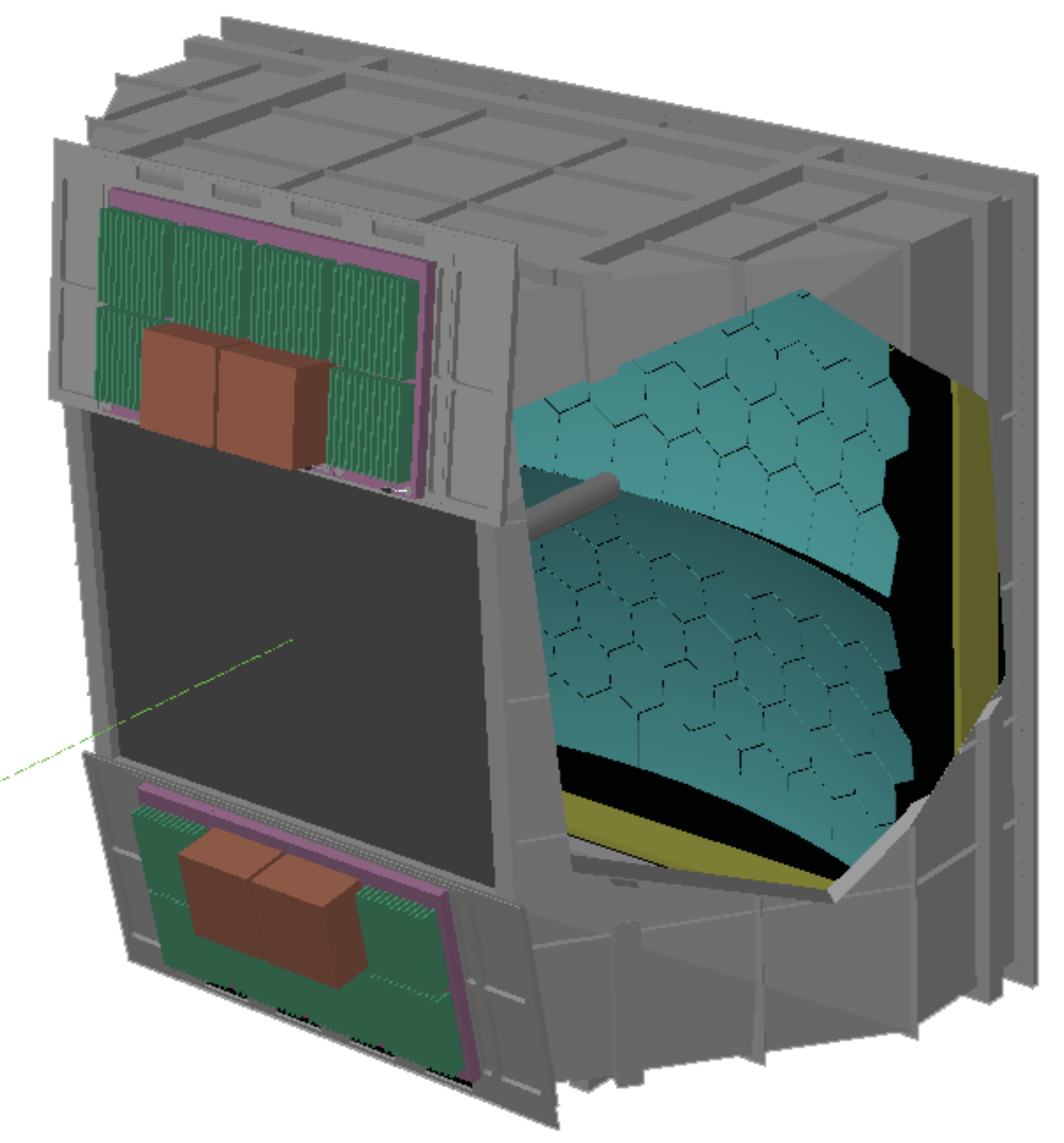}
\includegraphics[width=6cm]{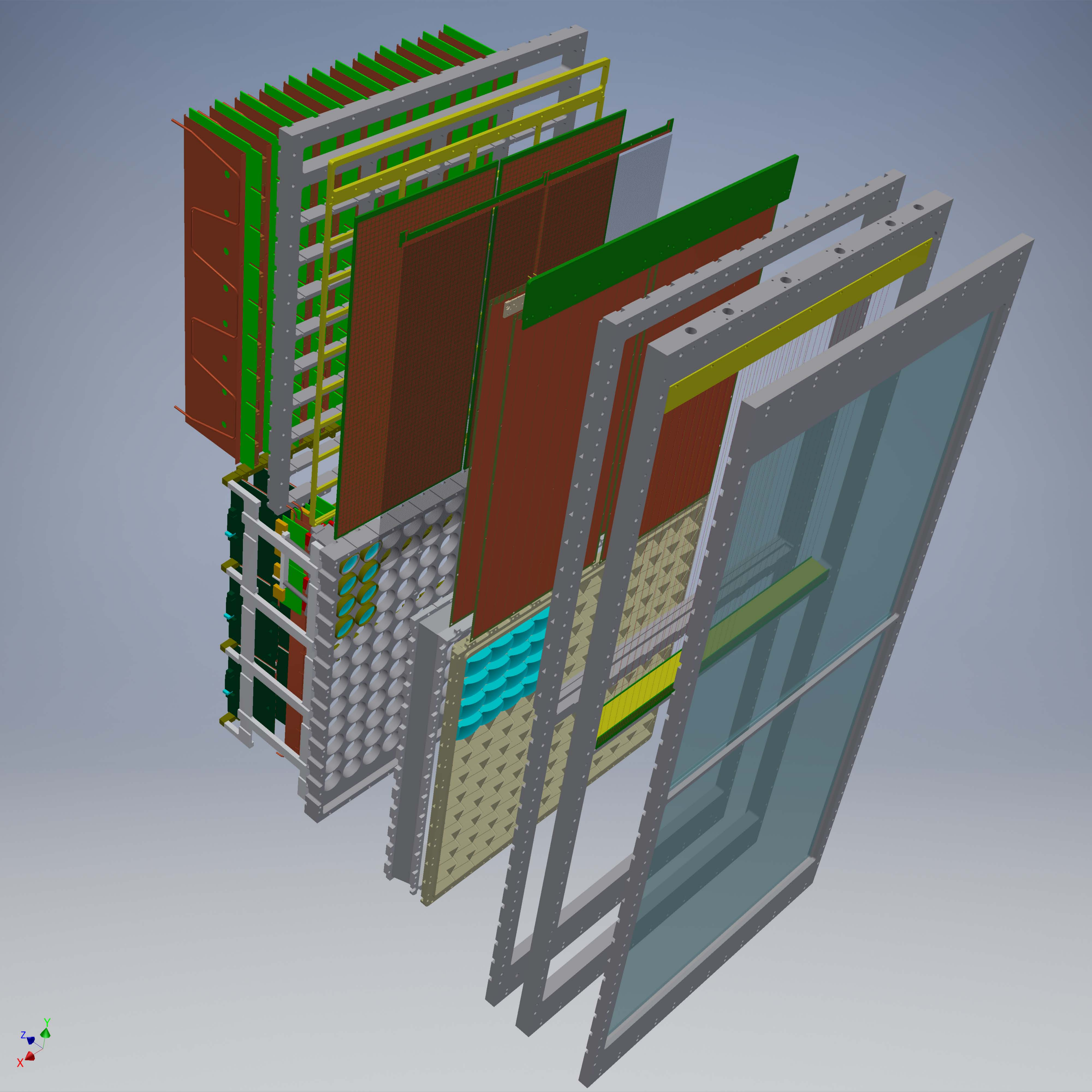}
\caption{Artistic view of the COMPASS RICH-1 detector (left) and exploded view of the new PDs (right)}
\label{rich}       
\end{figure}

COMPASS RICH-1 (Fig.~\ref{rich}, left) is a Ring Imaging Cherenkov counter with a 3~m long
gaseous C$_4$F$_{10}$ radiator, a 21~m$^2$ large focusing VUV~mirror surface and
Photon Detectors (PDs) covering a total active area of 5.5~m$^2$ with two technologies:
Multi-Anode PMTs coupled to individual fused silica lens telescopes
in the central region (25\% of the surface) and 
MWPCs equipped with CsI-coated photocathodes in the remaining surface.
\par
In spite of their good performance, MWPC-based PDs present intrinsic
limitations: aging (decrease of quantum
efficiency) after a few mC/cm$^2$ charge collection, feedback pulses with
a rate increasing at large gain values, long recovery time after occasional
discharge in the detector and long signal formation time.
The MWPCs have to be operated at low gain and present a
non-negligible detector memory and dead time. Among the eight MWPC's of
COMPASS RICH-1 four, located above and below the centre of the detector
show particularly critical performance~\cite{aging}.
The present upgrade aims at replacing them with novel MPGD-based PDs,
developed in a seven year-long dedicated R\&D programme~\cite{thgem-based-PD}:
the resulting detector architecture is a Hybrid MPGD
including two THick GEM (THGEM) multiplication
stages followed by a Micromegas.
\par
The main technical characteristics and the
production process of the MPGD-based PDs for COMPASS RICH-1 upgrade are
described in this article.
\section{The THGEM production and quality assessment}
\label{sec-1}
Each one of the four new chambers (Fig.~\ref{rich}, right) will host one of the
existing 600 $\times$ 600 mm$^2$ panel of Multi-Anode PMTs and lens telescopes
together with a new MPGD-based PD consisting of two identical hybrid modules
covering about 600 $\times$ 300 mm$^2$, arranged side by side.
\par
\begin{figure}
\centering
\begin{minipage}{.4\textwidth}
\centering
  \includegraphics[width=5.5cm,clip]{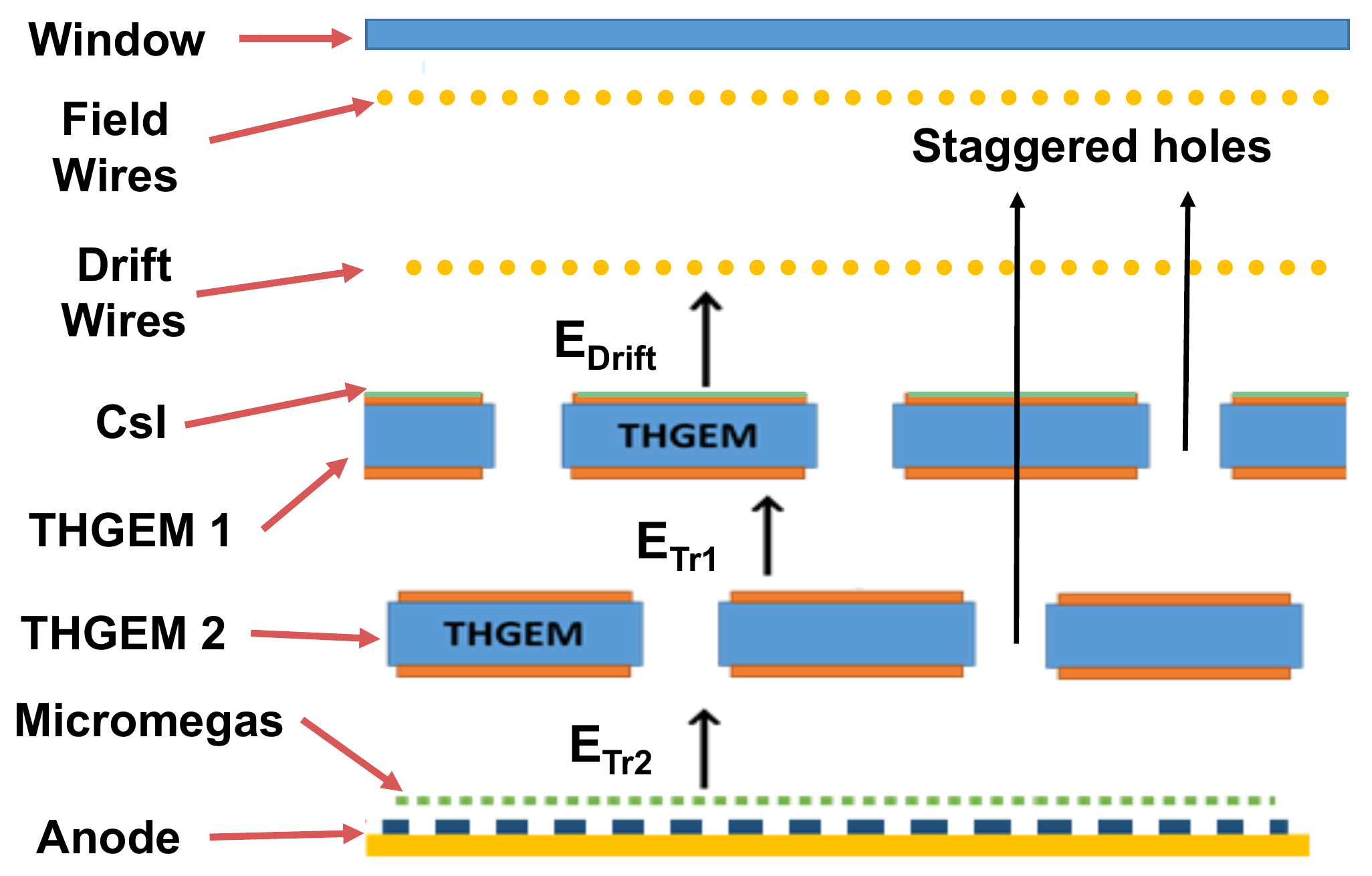}
  \caption{Schematic section of the hybrid module (not in scale).}
  \label{hybrid}
\end{minipage}%
\hfill
\begin{minipage}{.5\textwidth}
  \centering
  \includegraphics[width=7cm,clip]{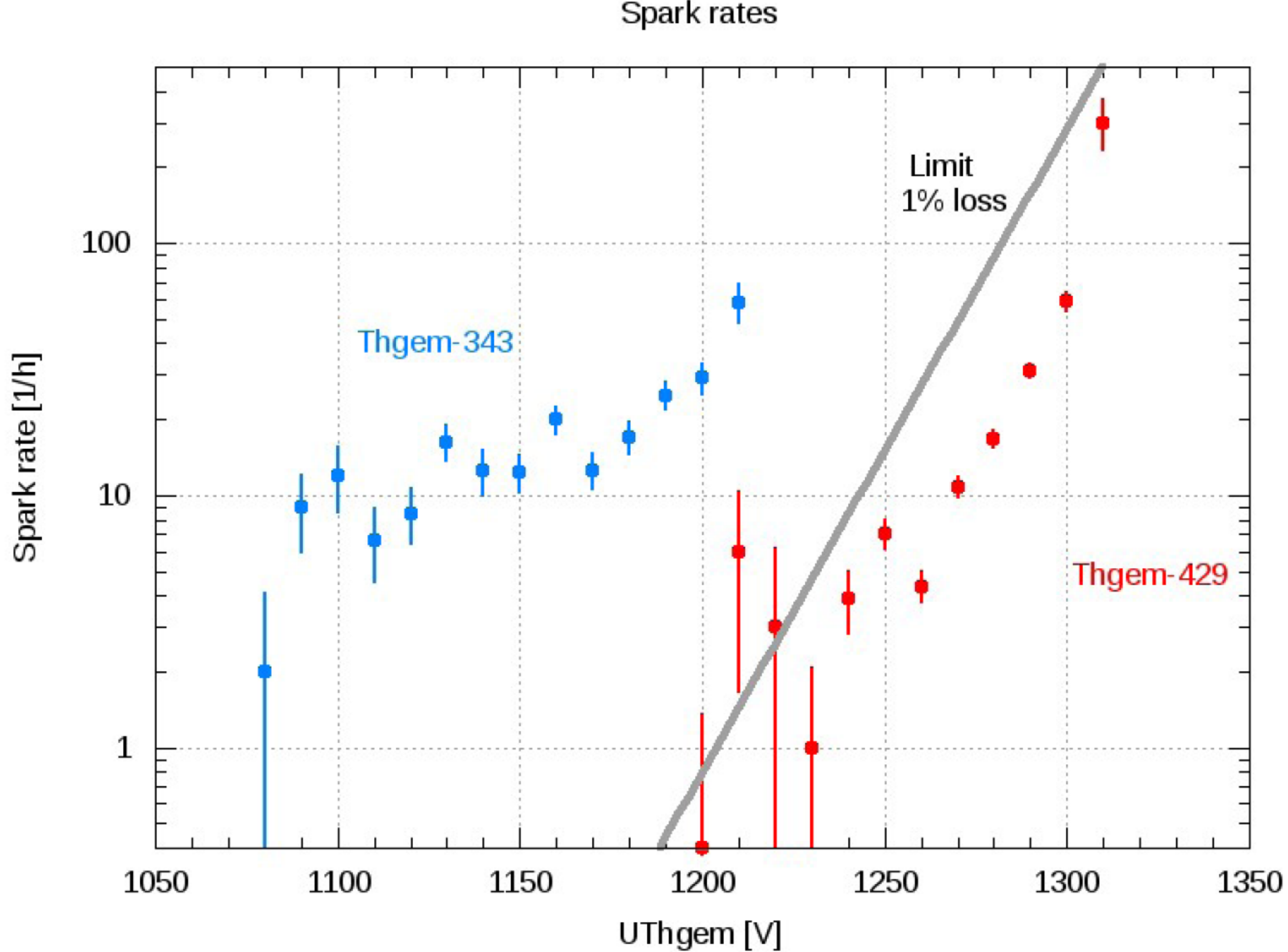}
  \caption{Rate of discharges in THGEMs as function of the bias voltage
	in Ar/CO$_2$ 70/30 gas mixture.}
  \label{discharge}
\end{minipage}

\end{figure}

The basic structure of the hybrid module (Fig.~\ref{hybrid}) 
consists of two layers of THGEMs, one Micromegas, and two planes of wires.
The top layer of the first THGEM (as seen from inside the RICH vessel) is coated
with a
CsI film and acts as a reflective photocathode for VUV photons.

\par
The THGEM PCBs
are all identical: they have an average thickness of 470 $\mu$m
(400 $\mu$m of fiberglass and 2 $\times$ 35 $\mu$m Cu), a length of 581 mm and a with of
287 mm. They have a regular hexagonal array of holes of 400 $\mu$m diameter, with a pitch of
800 $\mu$m. The holes are produced by mechanical drilling and have no rim.
The holes of the external border lines have 500 $\mu$m diameter, to avoid problems related
to the asymmetric field configuration at the electrode edges. 
On both top and bottom face the active area is divided in 12 sectors which
are 564 mm long and 23.3 mm wide (the two external ones are 17.9 mm wide), with a
separation of 0.7 mm between neighboring sectors.
Electrical connections are individually provided to each sector on one side of the THGEM.
A set of 24 holes, used as fixation points, guarantees the correct planar positioning
of the THGEM.
\par
The PCB raw material (halogen-free EM 370-5 from Elite Material Co, Ltd.) has been carefully
selected before the THGEM production phase: 50 foils (of 1245 mm $\times$ 1095 mm) have
been purchased, reduced to 800 mm $\times$ 800 mm by cutting out the borders which usually
present larger local thickness variations and mapped by means of a Mitutoyo EURO CA776
coordinate measuring machine
hosted in a thermally stabilized room.
The PCB foil was laying on the reference surface of the coordinate measuring machine,
kept flat by under-pressure while its top surface was mapped in a square pattern of points at
20 mm steps; the measurement was than repeated after reversing the foil and good consistency
between the two measurements was obtained. The reference surface
was regularly re-measured and the point by point difference between reference surface and
foil measurement provided the local foil thickness values. A typical thickness distribution
for a good foil presents an average value of 472 $\mu$m and a standard deviation of 2 $\mu$m.
Each PCB foil can host two raw THGEM PCBs, and a thickness uniformity quality factor
$\delta_{th}$ is evaluated for each surface of the PCB corresponding to a raw THGEM PCB.
The quality factor is defined as $\delta_{th} = (th_{max} - th_{min})/th_{min}$
where $th_{max}$ and $th_{min}$ are the maximum and minimum of the measured thickness values
in the active area of the THGEM. The best raw THGEM PCBs have $\delta_{th}~\leq$~2$\%$, and out of
a total of 100 raw THGEM PCBs, 49 have $\delta_{th} <$ 3$\%$, 38 have 3$\%$ $ \leq \delta_{th} <$
4$\%$ and 13 have $\delta_{th} \geq$ 4$\%$.
\par
After the mapping and selection of the raw material 42 THGEMs have been produced.
Individual foil labels and orientation marks allow to use the database of the local thickness
values in the later stages of the THGEM production, test and operation.
The mechanical drilling of the THGEM holes has been performed by ELTOS S.p.A. (Arezzo, Italy)
after the image transfer, development, etching and dry film stripping operations have been
completed. Up to 6 THGEMs can be drilled in parallel by using a multi-spindle machine
(Posalux 6000-LZ). To guarantee a good quality of the hole surface no more than 1000 holes
are drilled by a single tool. No electroplating process is applied and electrical continuity
between top and bottom paths for the voltage bias is provided by copper rivets where needed.
\par
A specific smoothing and cleaning procedure has been developed~\cite{polishing} and is
applied to the 42 THGEMs in the Trieste COMPASS Laboratory: it consists in a long,
careful polishing operation using fine-grain pumice powder and a manually guided
vibration-rotation machine, followed by cleaning using high pressure water and ultrasonic
bath in a basic solution (Sonica PCB detergent with PH=11), distilled water rinsing and
drying by blowing nitrogen and storing in a vacuum tank or nitrogen flushed box.
\par
For the quality assessment tests the THGEM is mounted in a box equipped with a pad-segmented anode
and a wire plane biased to provide a proper drift field; it is flushed with an Ar/CO$_2$ 70/30
gas mixture at 10 l/h flow and connected to a remotely controlled power supply 
programmed to provide from each electrode current monitoring and discharge counting.
Discharges are defined as monitored current values larger than 50 nA. The electrical stability test
program increases the bias voltage in steps of 10 V until a few discharges are registered; the bias
voltage is thus lowered and the test cycle restarts. Sparks are sometimes associated
to discharges and allow to visually identify the location of the discharge. Simultaneous discharges
often appear in neighboring sectors and also in distant sectors and even in different test boxes,
suggesting cosmic shower triggered events. A discharge rate below 1 event per hour is
considered acceptable and the maximum stable voltage is defined accordingly. If all sectors of the
THGEM have a maximum stable voltage larger than 1200 V (gain $\approx$ 3$\times$ the operational gain)
then the THGEM is validated for electrical stability.
\begin{figure}
\centering
\includegraphics[width=5.5cm]{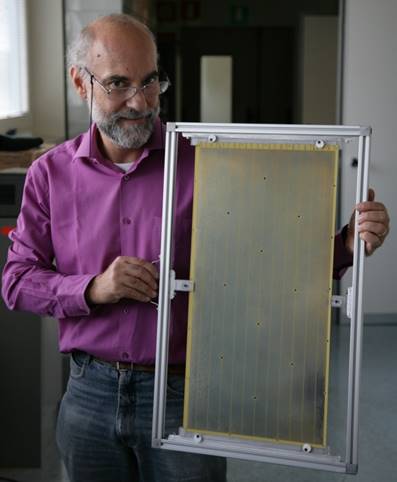}
\includegraphics[width=5cm,clip]{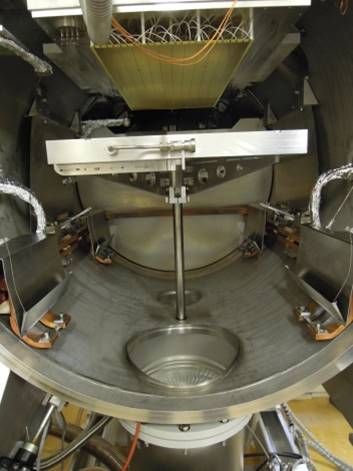}
\caption{A THGEM in its handling frame (left) and inside the CsI deposition plant (right).}
\label{discharge}       
\end{figure}
In  Fig.~\ref{discharge}, the rate of discharges for a typical validated THGEM (n.~429),
and for a rejected one (n.~343) are plotted as function of the bias voltage.
The line corresponding to the typical discharge rate for THGEMs at the validation threshold
limit is shown in gray.
The failure rate for this validation stage is about 30$\%$ but reapplying the polishing and cleaning
procedure allows to recover some of the THGEMs which fail the first electrical stability test.
\par
The gain uniformity of each THGEM has been measured for the first produced pieces by using a
$^{55}$Fe X-ray source sequentially positioned in 72 points and collecting the signal amplitude
spectra from the the set of anodic pads (covering about 20 cm$^2$) in front of the source position
via a CREMAT CR-110 preamplifier, a CANBERRA AFT Research Amplifier model 2025 and an AMPTEK 8000A
MCA. Typical standard deviation values range from 7$\%$ to 10$\%$ thanks to the strict thickness
selection criteria applied for the raw THGEM PCBs.
\par
For all THGEMs except the first ones a new procedure has been implemented for the gain uniformity
measurement: the X-ray source is an AMPTEK Mini-X with Au target and a Cu filter, providing
uniform illumination over the entire THGEM area at a rate of 5 kHz cm$^{-2}$, the readout being
performed by the RD51 Scalable Readout System (SRS) based on the APV-25 analogue readout chip. This
procedure is about 10 times faster and provides a detailed characterization of the local
gain variations.
\par 
The validated and characterized THGEMs are transported to CERN for the Ni-Au coating
process. A layer of about 5 $\mu$m thickness of chemically deposited nickel and about 0.5 $\mu$m
chemically deposited gold are used as protection against copper oxidation and as substrate for
the photoconverting CsI layer.
\par
The procedure developed by the RD26 Collaboration for the photocathode
production~\cite{braem} is followed,
whenever possible, for the CsI coating of the THGEMs faces to be used as reflective photocathodes.
After cleaning and preparation the THGEM to be coated is mounted in a dedicated gas tight box
with a two-pins electrical feed-through connection: the electrodes of the top layer of the THGEM
are connected together via one pin and the electrodes of the bottom layer are connected via the
other pin. The box is flushed with clean gas and used for THGEM storage but thanks to the
electrical connections the THGEM can be biased and its electrical strength checked before
and after the coating.
The 300 nm thick CsI layer is deposited by evaporation using the dedicated CERN plant, with the
typical conditions previously used for COMPASS and ALICE photocathode production (vavuum level
around 10$^{-6}$ mbar, T $\approx$ 60 $^{\circ}$C).
After the CsI evaporation a measurement of photocurrent is performed by illuminating with VUV
light from a D$_2$ lamp a spot of about 1 cm$^2$ and recording the ratio between the THGEM
photocurrent and the current from a reference PMT sensitive in the UV; the measurement is
performed at 60 different points to map the response uniformity and it is repeated after
lowering the temperature to $\approx$ 25 $^{\circ}$C, before extracting the coated THGEM
from the evaporation plant.
\section{The Micromegas and the other components}
\label{sec-2}
\par
COMPASS Micromegas are produced at CERN using the bulk technology (standard photolithographic
lamination technique) to fix a stainless steel woven square micromesh
(18 $\mu$m diameter wire, 63 $\mu$m pitch, tensioned at 15 N) onto an anode PCB specifically
designed for COMPASS RICH-1 Micromegas. The anode PCB is 3.2 mm thick and has four copper layers:
the anode layer is segmented in square pads of 8.0 mm pitch (with 0.5 mm inter-pad distance
and pads of 7.5$\times$7.5 mm$^2$ Cu with 0.3 mm radius at the corners); a second layer,
with the same segmentation,
sits inside the PCB at 70 $\mu$m from the anode, providing a capacitive coupling of $\approx$
40 pF between each anode pad an its facing (readout) pad, a value about 10 times larger
than the capacitance between an anode pad and the micro-mesh; two layers are needed to route
the paths from each readout pad to the corresponding pin of the connector for the front-end
electronics and from each anode pad (trough a hole at the center of the readout pad)
to the pin of the connector distributing the positive high voltage bias. Individual resistors
of 470 M Ohm are separating each anode pad from the high voltage power supply, to reduce
the effects of a local discharge and let the non discharging pads unaffected by the discharge
and recharge processes.
The uniformity of the Micromegas amplification gap (128 $\mu$m nominal distance) is guaranteed
by a square array of cylindrical micro-pillars with 300 $\mu$m diameter and 2 mm pitch
(16 per pad) obtained from the photo-imageable polyimide film.
A 5.5 mm wide coverlay border surrounding the detector's active area helps holding the micromesh
and 24 circular (6 mm diameter) coverlay spots are prepared as bases for the THGEM fixation pillars.
At the end of the production process the Micormegas is cut at its final 283$\times$586 mm$^2$
dimension, readout connectors and bias resistor connectors are mounted and two identical
Micromegas are glued onto the detector holder, side by side. The detector holder consists
of an aluminum frame machined (by electro-erosion) from a 30 mm thick plate to provide stiffness,
planarity and specific grooves for the insertion of the front-end boards, glued to a frame
made of isolating material to avoid parasitic capacitance and machined to host the Micormegas. 
Strong Micromegas grounding is provided by embedding in conductive glue a large surface
(5$\times$260 mm$^2$) of each micromesh, on the side which is not in the active area of the
final detector. 
\par
The gain uniformity of the Micromegas has been studied using a prototype detector,
an Ar/CO$_2$ 70/30 gas mixture and a $^{55}$Fe source:
for an average gain of about 3000 the the local gain variations typically show
a standard deviation of 5$\%$.


\par
The micromesh, being the most extended unique electrode in the photon detector, is always kept at
ground potential and in case of a discharge in the Micromegas the potential of the anode
pad where the discharge takes place completely drops. To study the influence of such events on
the neighboring pads, discharges have been systematically generated on a specific pad (by biasing
it with an anomalously large potential) and the adjacent pads have been biased using for each pad a
different high voltage channel from the power supply.
The probability of correlated (within $\pm$ 100 ns)
discharges turned out to be negligible (no discharge cross-talk) when the adjacent pad was operated
at normal voltage values, while raising the bias voltage of a neighboring pad led to more and more
frequent correlated discharges, with a probability up to 0.7 when the same anomalous bias was applied,
with a large majority of correlated events having a time difference $\leq$ 10 ns.
\par
After gluing the 48 THGEM pillars (24 for each Micromegas, M3 threaded, made of PEEK,) on the prepared
coverlay bases and the high voltage distribution PCBs on the detector frame the hybrid PD is
assembled: the gap between the Micromesh and the closer THGEM is 5 mm wide while the distance between
the THGEMs is 3 mm. The THGEMs are mounted in a staggered holes configuration (by a $\approx$ 462
$\mu$m displacement with respect to the aligned holes configuration) in order to favor the spread of
the avalanche charges.
\par
The electrostatic field above the photocathode is defined by two planes of wires
(100 $\mu$m diameter, Cu-Be alloy with a flash of Ni and Au), 60 cm long,
with a pitch of 4 mm and guard wires of 200 $\mu$m diameter at the edges.
The wires are manually soldered on half of a large (645 $\times$ 1400 mm$^2$) mechanical
frame which is also used, together with a similar size spacing frame, as holder
of the existing Multi-Anode PMT based photon detectors. 
Special electrodes, embedded in the isolating material protections of the chamber
frames are used to shape the field at the edges of the active volume, preventing
large field values on any of the electrodes in the gas volume: the optimization of
the final geometry required detailed numerical calculations of the fields taking
into account all electrodes and their real shapes.
\par
The preparation and mounting of the wire and spacing frames too requires several
different steps since the overall mechanical structure of the detector is rather complex
(see Fig.~\ref{rich}, right) and the procedures for assembling and installation
of the new PDs are quite elaborated.
\par
The high voltage distribution system provides independent bias values to each
half of a THGEM face (6 sectors are connected to one HV channel via a resistive splitter
with protection diodes to prevent HV drop cross-talk between sectors).
Specific switch-on and safe-mode procedures prevent anomalous bias values to be applied
to the THGEMs. 
The effective values of the applied bias will be continuously corrected to counteract
the effects of environmental variables (p,T) variation:
a gain stability in the order of $\pm$10$\%$ is expected, thanks to these corrections.
\par
The signal from the readout pads are collected by front-end electronic cards~\cite{APV}
designed for COMPASS RICH-1: they host four APV25-S1 chips, each reading 108 pads.
Three front-end cards are connected to a 10-bit flash ADC digitizer board equipped
with a FPGA performing on-line zero suppression. A cooling system using under-pressure
water flow assures efficient removal of the heat produced by the readout.
\par
The new hybrid PDs will be tested and installed on COMPASS RICH-1 before the 2016 run.

%
%
%
\section{Conclusions}
\label{concl}
COMPASS RICH-1 will operate during the physics run of 2016 in an upgraded version with four new
Photon Detectors based on hybrid THGEM + Micromegas technology. The expected high performance of
the new PDs will allow to meet the strict requirements of the new COMPASS physics measurements.
The production of all elements of the new PDs is almost complete and the procedures for test
and validation of the components have been defined and applied.
The recent developments of the field confirm the great potentialities of MPGD-based PDs and
offer promising perspectives for their future applications.

%
%

\end{document}